\documentclass[11pt,a4paper]{article}
\usepackage[T1]{fontenc}
\usepackage[utf8]{inputenc}
\usepackage{authblk}
\usepackage{graphics,graphicx,dcolumn,bm,fleqn,epic,eepic,float}
\usepackage{amssymb,amsmath,multirow,rotate,color,float}
\usepackage{graphicx} 
\usepackage{color}
\definecolor{red}{rgb}{1,0,0}
\definecolor{green}{rgb}{0,0.5,0}
\definecolor{blue}{rgb}{0,0,1}
\usepackage{tabularx}

\title{The effect of the number of states on the validity of credit ratings}

\author[1,2]{Pedro Lencastre}
\author[3]{Frank Raischel}
\author[4]{Pedro G.~Lind}
\affil[1]{ISCTE-IUL, Av.~For\c{c}as Armadas, 1649-026 Lisboa, Portugal\\
  (e-mail: {\tt pedro.lencastre.silva@gmail.com})}
\affil[2]{Mathematical Department, FCUL, University of Lisbon, 1749-016 Lisbon, 
  Portugal}
\affil[3]{Instituto Dom Luiz, University of Lisbon, 
  1749-016 Lisbon, Portugal}
\affil[4]{Instituto Dom Luiz, University of Lisbon, 
  1749-016 Lisbon, Portugal}
\begin{document}

\maketitle             

\begin{abstract}
We explicitly test if the reliability of credit ratings depends on the total
number of admissible states. We analyse open access credit rating data and show that the effect of the number of states in the dynamical properties of ratings change with time, thus giving supportive evidence that the ideal number of admissible states changes with time.
We use matrix estimation methods that explicitly assume the hypothesis needed for the process to be a valid rating process. By comparing with the likelihood maximization method of matrix estimation, we quantify the ''likelihood-loss'' of assuming that the process is a well grounded rating process.
\end{abstract}

\section{Motivation and Scope}

Credit ratings are a popular tool to evaluate credit risk and calculate the 
capital adequacy requirements for banks \cite{peura2004simulation}. However, there is usually no explicit criteria to determine the number of states a 
rating scale should have. This is true for both the credit ratings published 
by the credit rating agencies and for the internal ratings used by 
banks \cite{grunert2005role}.
It has been argued that the validity of the rating process as a measure of 
credit risk depends on it being Markov and 
time continuous \cite{lencastre2014credit}. If ratings follow criteria 
based on financial and economic variables, which are time continuous, then 
they should themselves be time continuous. If this fails, then it is possible 
that ratings are biased by concerns other than financial and economical ones. 

In this paper, we discuss the possibility that the dynamical properties of 
the process change when we change the number of states. If the number of 
states is artificially large, there might be additional constrains to make 
the process stable or to ascribe the ratings. In turn, these additional constrains can affect the validity 
of the usual assumption that ratings are a time continuous process. On the other 
hand, if there are too few states, then it might be impossible to distinguish 
two issuers with the same rating based on their intrinsic credit risk. 
Furthermore, in this case, it is possible that the historical data of 
previous ratings might be a criterion to determine the different levels of 
credit risk within the same rating state, meaning that the process is not 
Markov. 
\begin{figure}[htb]
  \centering
  \includegraphics[width=0.48 \linewidth]{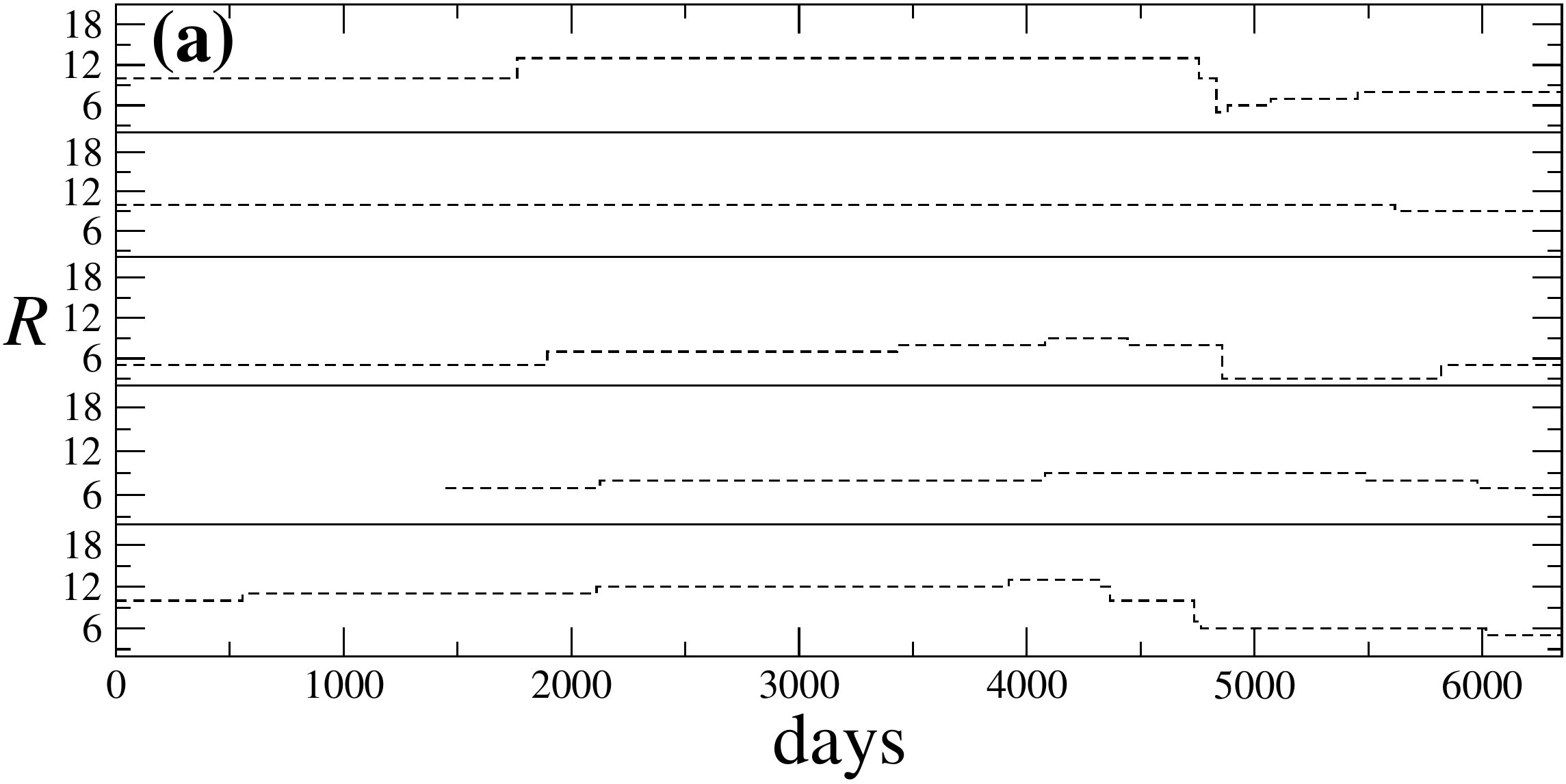}
  \includegraphics[width=0.23 \linewidth]{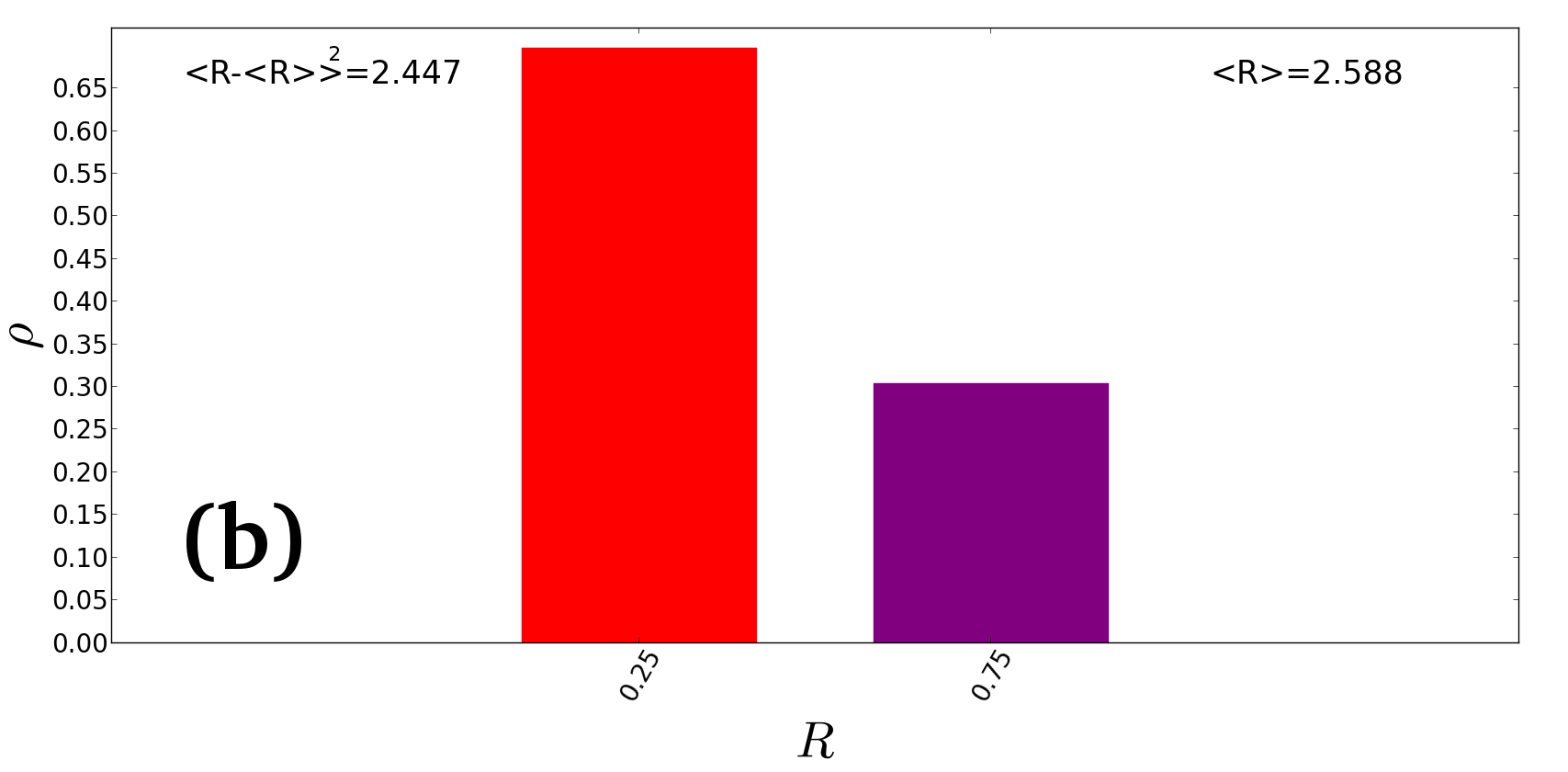}%
  \includegraphics[width=0.23 \linewidth]{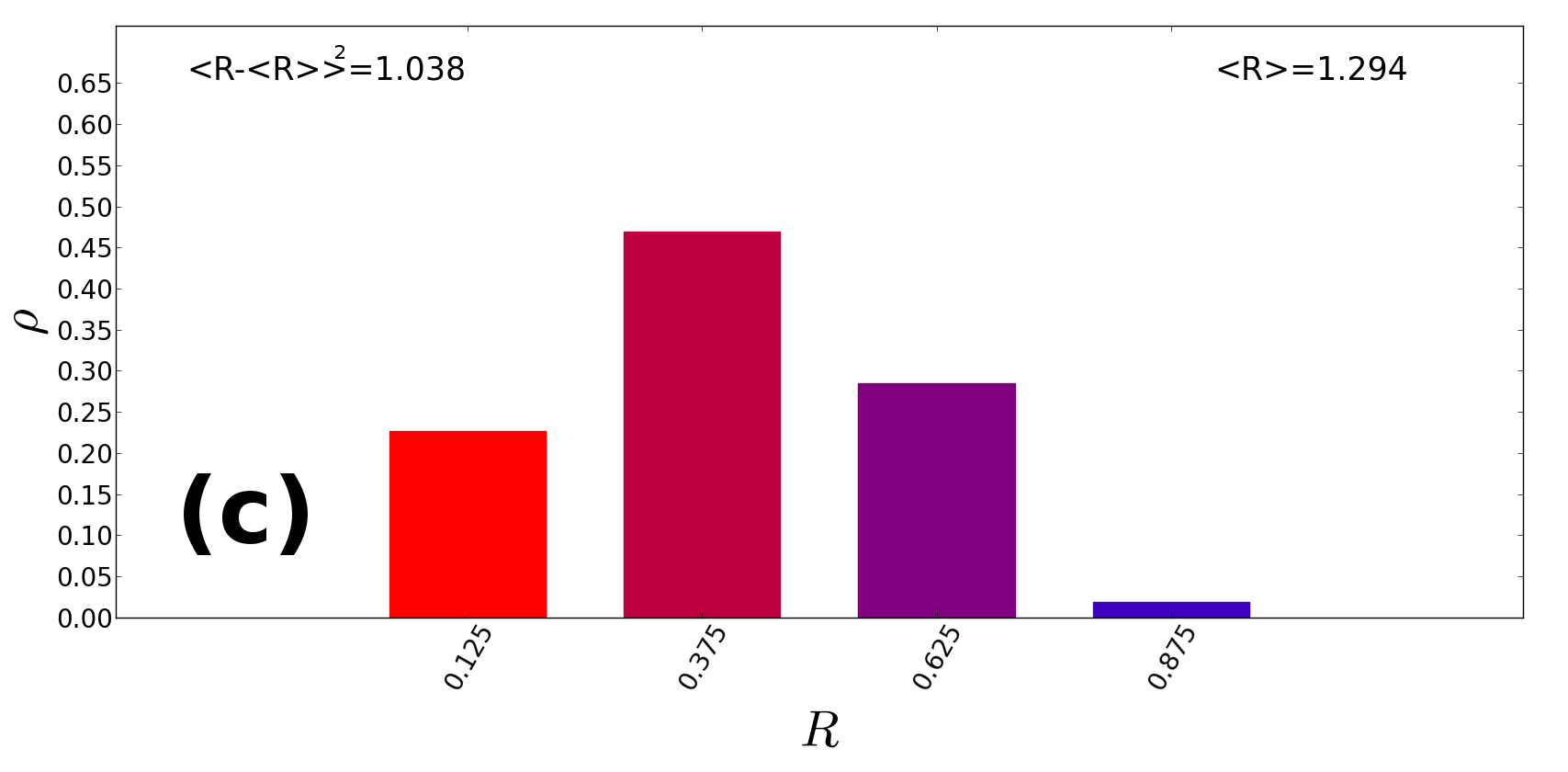}%
  \includegraphics[width=0.23 \linewidth]{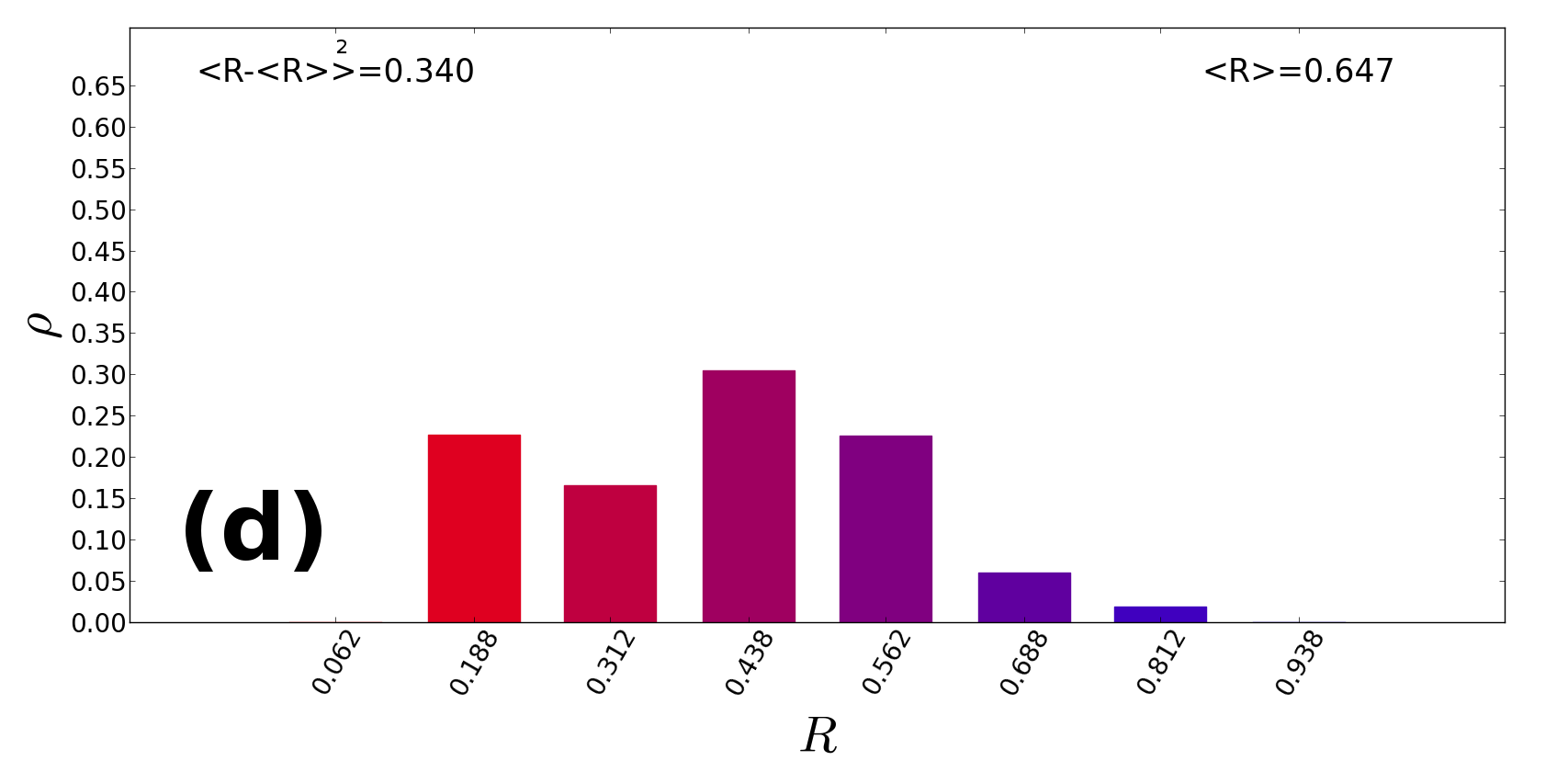}%
  \includegraphics[width=0.23 \linewidth]{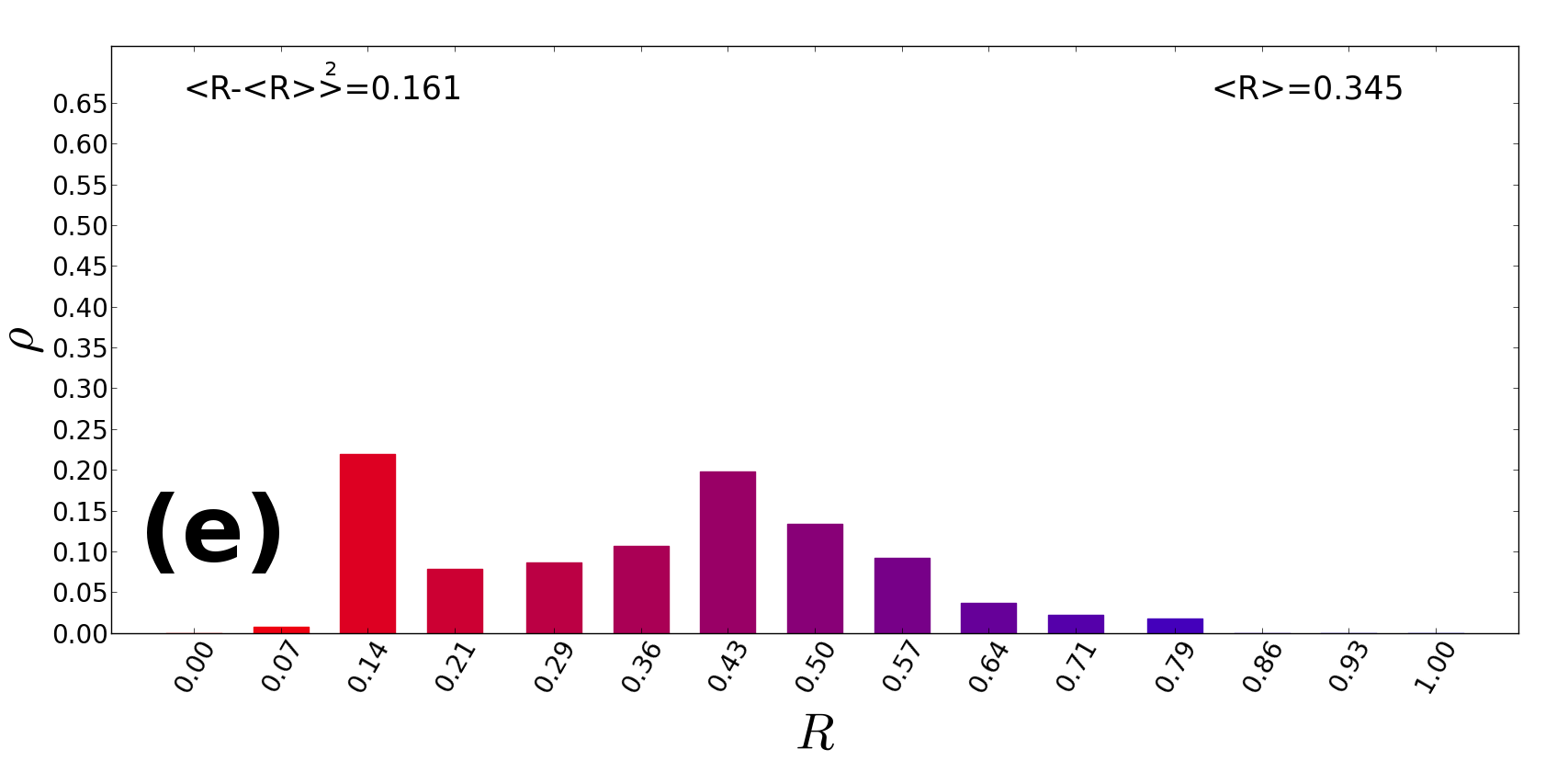}
  \caption{\protect
  {\bf (a)}
  Time series of the credit ratings of $10$ individual entities from the time 
  when they are first rated until the present. 
  {\bf (b-e)} Histogram of rating frequencies 
  in the first of January of 2010 for different number of states 
  $n_s= \{2,4,8,15 \}$.}
  \label{fig01}
\end{figure}
By using different techniques to estimate a transition matrix from a finite 
sample of data \cite{Lando2002}, one can evaluate the dependency of the 
dynamical properties of ratings with the number of states. 
Below, we compare transition matrices calculated under different assumptions 
and show that the quality of the time continuity and Markov assumptions 
changes considerably in time. 
We start, in Sec.~\ref{sec:empirical_data}, by describing the empirical 
data collected from Moody's.
In Sec.~\ref{sec:theory} we present the theoretical background
for estimating transition matrices and in Sec.~\ref{sec:findings} 
we describe how to test the validity of both the existence of a generator
and Markov assumptions.
Section \ref{sec:conclusions} concludes the paper and presents some
discussion of our results in the light of finance rating procedures.

\section{Data description}
\label{sec:empirical_data} 

The time series used in this paper was reconstructed by 
us\cite{lencastre2014credit} from data provided by Moodys, an influential 
credit rating agency, and publicly available in compliance with Rule 
17g-2(d)(3) of US.~SEC regulations \cite{Database}. 
Our data sample is the set of rating histories from the European banks,
with a sample frequency of one day, starting in January 1st 2007 and ending 
in January 1st of 2013. 

The rating class considered for our analysis is the so-called 
\textit{Banking Financial Strength}\cite{moody2004definitions}, 
a measure of a bank's intrinsic credit risk including factors such as
franchise value, and business and asset 
diversification \cite{moody2004definitions} and excluding external 
factors, such as government support.
In the Financial Strength rating class there are 15 credit rating grades, 
represented by letters from "A" to "E" with the two possible extra suffixes 
"+" and "-". 

Figure \ref{fig01}a shows several examples of individual rating 
histories for each entity. It can be seen that rating changes occur infrequently, with an average of $0.43$ transitions each year. 
We further label each state 
with a number ranging from $\frac{1}{2n}$ (default) to $1 - \frac{1}{2n}$ (highest rating), where $n$, with $n >1$, represents the number of accessible states. 

In Figs.\ref{fig01}b-e we plot the histogram of the ratings
$R$ on January 1st 2010, for $2,4,8$ and $15$ states respectively.
The histograms were constructed from the original $15$-state histogram 
by merging pairs of adjacent states into one single state.


\begin{figure}[htb]
\centering
\includegraphics[width=0.8 \linewidth]{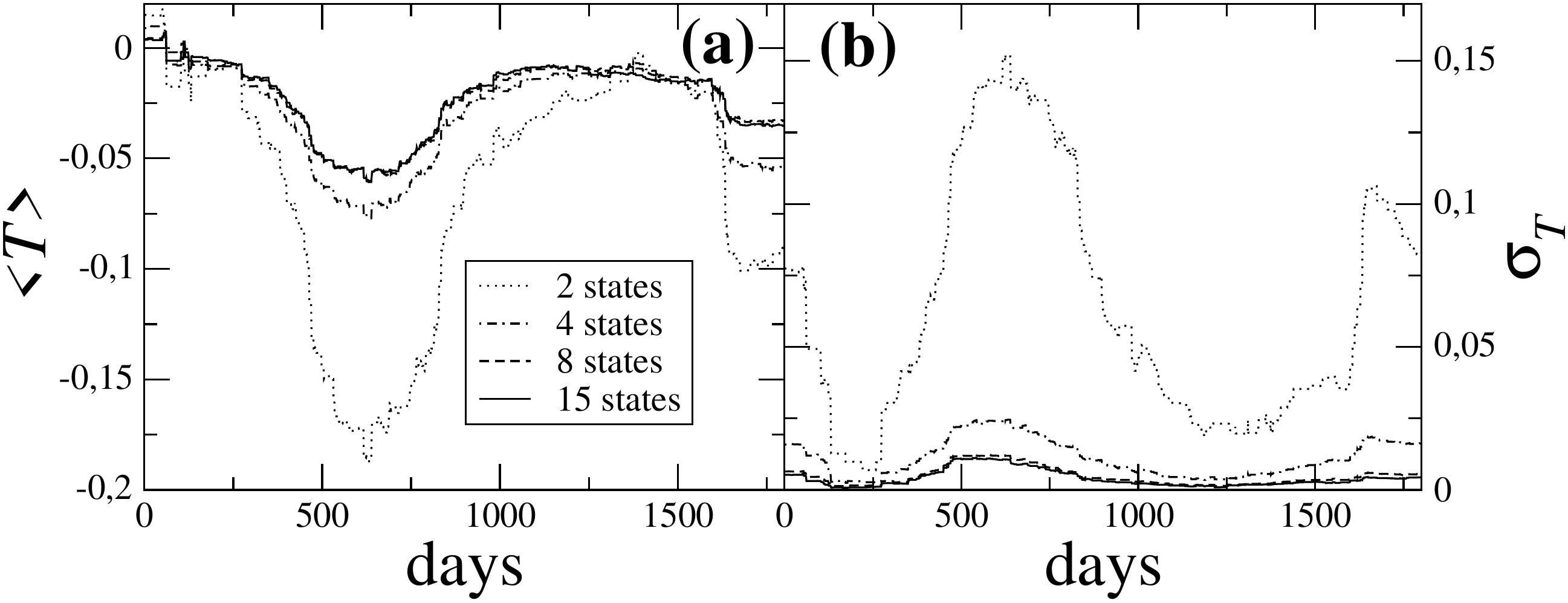}
\caption{\protect
         Comparing the sensitivity of the rating increment evolution
         with the number of admissible states. For $2,4,8$ and $15$
         admissible states, one plots
         {\bf (a)} the mean $\langle T(t,\tau)\rangle$ of the ratings 
                   increments, 
         {\bf (b)} the standard deviation $\sigma_T(t,\tau)$,
         both for $\tau=1$ year.}
\label{fig02}
\end{figure}

We define the rating increments as
\begin{equation}
T_i(t,\tau) = R_i(t) - R_i(t-\tau) .
\label{transition_process}
\end{equation}
When $T_i(t)>0$ (resp.~$<0$) it means that bank $i$ saw its rating 
increased (resp.~decreased) during the last $\tau$ period of time.
Unless stated otherwise we will use always $\tau = 1$ year.

In Fig.~\ref{fig02} one sees that, although the first two statistical 
moments of $T(t,\tau)$ grow when we use less states, they
have always a similar shape whatever number of admissible states
is used.
\section{Estimating rating transition matrices}
\label{sec:theory}

There are three methods of extracting a transition matrix directly from a 
time series \cite{jafry2004measurement}.

One is simply the normalization of the transition number matrix $\mathbf{N}$, 
where each entry $N(t,\tau)_{i,j}$ gives the number of transition in the desired 
time interval $[t-\tau,t]$. 
The transition matrix is then obtained by normalizing the row-sums to one:
\begin{equation}
(\mathbf{T}(t,\tau))_{i,j} = \frac{(N(t,\tau))_{i,j}}{\sum_j (N(t,\tau)_{i,j}}  .
\end{equation} 
This estimation method is called the Cohort Method\cite{Lando2002} and is 
the default method. 

Another method consists in estimating the generator matrix \cite{Israel2001},\cite{Davies2010}
$\mathbf{Q}$ directly from empirical 
data\cite{Lando2002,Charitos2008computing}. 
The off-diagonal elements are empirically estimated as
\begin{equation}
(Q(t,\tau))_{i,j} = \frac{(N(t,\tau))_{i,j}}{ \int_{t - \tau}^{t} N_i(t\prime) dt\prime} ,
\label{empirical_homogeneous_generator}
\end{equation}
where the number $N_i(t\prime)$ stands for the 
number of entities in  state $i$ at  moment $t$.
The diagonal elements are calculated by forcing the row-sums of $\mathbf{Q}$ 
to be zero. 
We then calculate the transition matrix as $\mathbf{T}^{\prime}(t,\tau) = e^{\mathbf{Q}(t,\tau)}$. 
This estimation method is valid if and only if the underlying process is 
time-continuous, Markov and time-homogeneous\cite{Lando2002}. 

Finally, the third method consists in using the Chapman-Kolmogorov
Equation\cite{risken}, which holds for any Markov process, given
by
\begin{equation}
\bar{\mathbf{T}}(t,\tau) = \prod_{n=1}^{k}\mathbf{T}_{t-(k-i)\tau,\tau \prime } ,
\label{Chapman-Kolm}
\end{equation}
where $\tau \prime = \tau / k$. This corresponds to the multiplication of matrices with a smaller non-overlapping time-window of size $\tau \prime$.

\section{Comparing different estimations}
\label{sec:findings}

In this work we want to compare all three estimations,
$\mathbf{T}$, $\mathbf{T}^{\prime}$ and $\bar{\mathbf{T}}$
and quantify the difference between them. 
To compare them, we will use a so-called likelihood difference
between the transition matrix of the default method $\mathbf{T}$
and the other two estimations:
\begin{equation} \label{eq:log-likelihood}
d(\mathbf{T},\mathbf{T}_{other}) =
\frac { \sum_{i,j}  (N(t,\tau))_{i,j} \left(  \frac{\log{(T(t,\tau)_{i,j})}}{\log{((\mathbf{T}(t,\tau)_{other})_{i,j})}} \right) }{ \sum_{i,j} (N(t,\tau))_{i,j}} \,
\end{equation} 
where $\mathbf{T}_{other}$ represents either
the matrix $\mathbf{T}^{\prime}$ of the generator matrix estimation method or the matrix
$\bar{\mathbf{T}}$ in the Chapman-Kolmogorov estimation method. 
This norm is particularly useful when one of the matrices maximizes of the 
likelihood function, which is the case for the cohort estimated matrix \cite{Lando2002}. 
This way, the distance $d(\mathbf{T}_{other} , \mathbf{T})$ becomes an adequate measure 
of the "likelihood loss" by choosing a different estimation method.
\begin{figure}[tb]
  \centering
  \includegraphics[width=0.95 \linewidth]{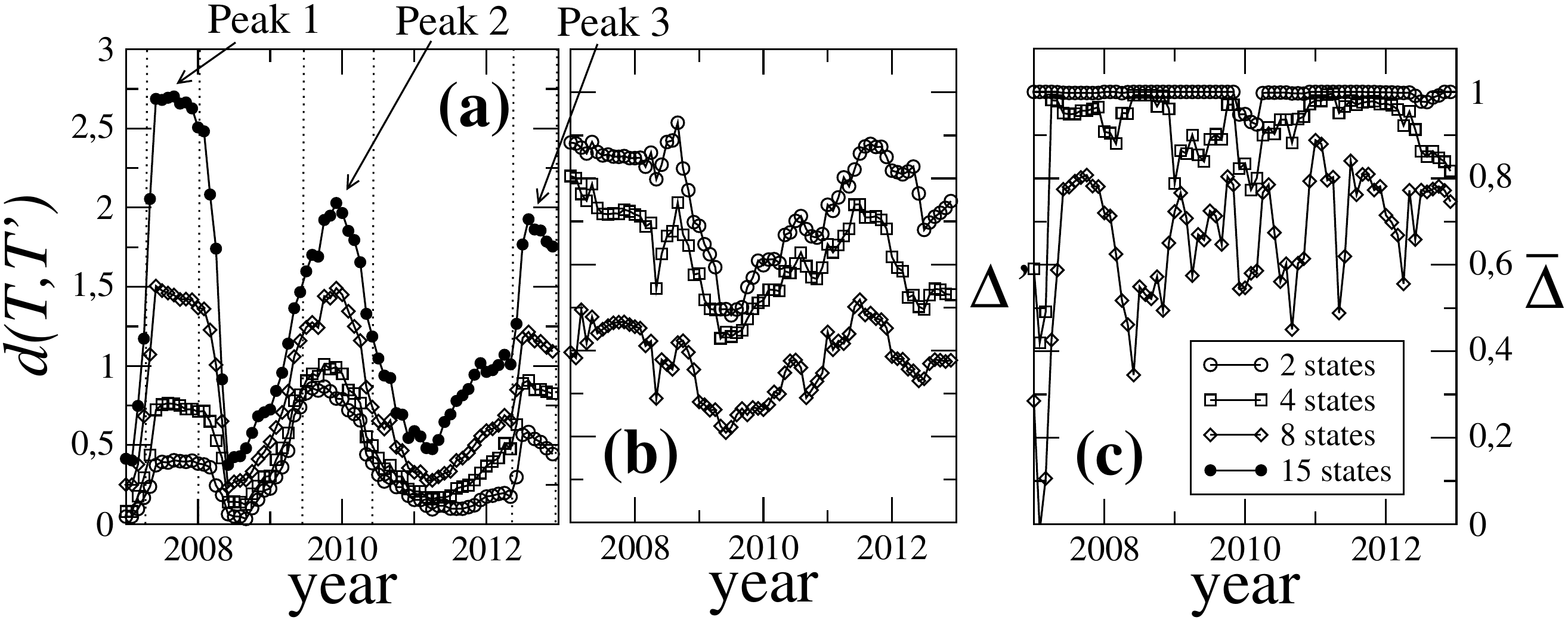}
  \caption{\protect
  {\bf (a)} Comparison between $\mathbf{T}$ and $\mathbf{T}\prime$, 
  calculated over a time-window of one year, 
  using the log-likelihood distance of the Eq.~\ref{eq:log-likelihood}. 
  {\bf (b)} Comparison between the likelihood distance with $15$ states and
  a smaller number of states.
  {\bf (c)} The same as (b) for the transition matrix $\overline{\mathbf{T}}$
  computed from the Chapman-Kolmogorov equation.}
  \label{fig03}
\end{figure}

We compare $\mathbf{T}$ with $\mathbf{T}\prime$, and $\overline{\mathbf{T}}$ 
with a fixed time-window of one year, and plot $d(\mathbf{T},\mathbf{T}\prime)(t)$ 
and $d(\mathbf{T},\overline{\mathbf{T}})(t)$ in Fig. ~\ref{fig03}.
If the process is time-homogeneous, Markov and time-continuous, as one expects from 
a rating process \cite{lencastre2014credit}, then the difference between $\mathbf{T}$, 
$\mathbf{T}\prime$ and $\overline{\mathbf{T}}$ should be white noise 
resulting from dealing with a finite sample. 

However, 
when too few states are considered, e.g.~only two states (good and bad), it might be 
possible to distinguish entities in the same state due to their history, making ratings 
not Markov.
In general we expect 
the system to lose the Markov property when the ``width'' of each rating grade, 
a percentile of the maximum financial health, is bigger than the uncertainty associated 
to the uncertainty characteristic of the average entity's financial health. This might be counter-balanced by the fact that when we reduce the number of states we also lose information about the system's past. 
Nonetheless, when we lower the number of states, the ``pure'' white noise of dealing with a 
finite sample of data becomes lower, but the difference between matrices might become 
larger due to the fact that the process might not be Markov.


To see if the process is time-continuous, Markov, and time-homogeneous, we 
measure how correct it is to assume that a generator exists through
the difference $d(\mathbf{T}, \mathbf{T}\prime)$. Results are shown in 
Fig.~\ref{fig03}a. In Fig.~\ref{fig03}b one plots the normalized difference
between the distances for the standard $15$ states and the fewer states cases:
\begin{equation}
\Delta(\mathbf{T},\mathbf{T}_{other}) = \frac{d(\mathbf{T},\mathbf{T}_{other}^{(15)})-d(\mathbf{T},\mathbf{T}_{other}^{(n)})}{d(\mathbf{T},\mathbf{T}_{other}^{(15)})} ,
\label{delta}
\end{equation}
with $\mathbf{T}_{other} = \mathbf{T}\prime$ and with $n=2,4$ and $8$. As a notation simplification, we use $\Delta\prime = \Delta(\mathbf{T},\mathbf{T}\prime)$ and $\bar{\Delta} = \Delta(\mathbf{T},	\bar{\mathbf{T}})$.

When testing the validity of the assumption of the existence of a generator, it possible to observe that the shape of 
$d(\mathbf{T},\mathbf{T}\prime)$ 
preserves most features when we change the number of states, although the 
scale of the different graphics change.
In fact, as can be seen from Fig.~\ref{fig03}b, all three peaks in 2007, 2009 and 2012 remain visible 
and occur at approximately the same time and have the same duration. 
Their relative strength, however, is inverted. With $15$ rating classes, the peak in 2007 is stronger than the other 
two, equally strong  peaks in 2009 and 2012. The relative strength of the 
three peaks is somehow equal using the reduced sets of 8 and 4 states, 
whereas for two states the 2009 peak becomes strongest. 


To test when the process is Markov we check when does the Chapman-Kolmogorov 
Equation (\ref{Chapman-Kolm}) hold, and in Fig.~\ref{fig03}c one plots a quantity $\bar{\Delta}$ using Eq.~(\ref{delta}) but this time for using the matrices $\mathbf{T}$ and $\bar{\mathbf{T}}$.
The peaks of 
$d(\mathbf{T},\overline{\mathbf{T}})$ occur now at the same instants as the peaks 
of $d(\mathbf{T}, \mathbf{T}\prime)$, namely at the beginning of 2007, 2009 and 
2012. 

Reducing the number of states preserves the shape of the graphic, and the years 
at which a greater deviation occurs, but greatly diminishes the overall scale 
and difference between $\mathbf{T}$ and $\overline{\mathbf{T}}$. In the limit of two states, almost all information is lost.



\section{Conclusions}
\label{sec:conclusions}

We have used publicly available Moody's time series of credit ratings and
studied simple ways to compute the validity of the usual assumptions, regarding credit rating time series, as a function of the number of rating
states. 

In general, we have presented evidence that the accuracy of the 
time-continuous Markov time homogeneous assumptions vary not only
in time but also with the number of states. 
Consequently, the choice of how many admissible states one should
use to categorize a sample of rated entities should not be
fixed in large time-spans.

Moreover, we also found that during periods when the process shows to be
non-Markovian or during which no proper generator exists the choice of
the number of admissible states is of importance, since the non-Markovianity
and the unreliability of a generator is less pronounced when one chooses
a number of admissible states smaller than the standard $15$ states.

Using empirical data for which $15$ states were prescribed it is not possible
to extend this work to consider a number of admissible states larger than
$15$. One may conjecture that, the choice of $15$ is better than a smaller
number for detecting deviations from Markov time-continuous processes, but
whether or not this is an optimal choice is an open question for future
investigations.

\section*{Acknowledgments}

$ \quad \, $The authors thank 
Funda\c{c}\~ao para a Ci\^encia e a Tecnologia 
for financial support 
under PEst-OE/FIS/UI0618/2011, PEst-OE/MAT/UI0152/2011, 
 FCOMP-01-0124-FEDER-016080, SFRH/BPD/65427/2009 (FR).
This work is part of a bilateral cooperation DRI/DAAD/1208/2013 
supported by FCT and Deutscher Akademischer Auslandsdienst (DAAD).
PL thanks Global Association of Risks Professionals (GARP)
for the ``Spring 2014 GARP Research Fellowship''.

\section*{References}

\bibliographystyle{plain}
\bibliography{MatrixBib}

\end{document}